\documentclass[prc,twocolumn,showpacs]{revtex4}
\usepackage{graphicx}
\usepackage{dcolumn}
\usepackage{bm}

\begin{document}

\title{Stellar weak decay rates in neutron-deficient medium-mass nuclei}

\author{P. Sarriguren}
\affiliation{Instituto de Estructura de la Materia, CSIC, Serrano
123, E-28006 Madrid, Spain}



\begin{abstract} 

Weak decay rates under stellar density and temperature conditions holding
at the rapid proton capture process are studied in neutron-deficient 
medium-mass waiting point nuclei extending from Ni up to Sn.  
Neighboring isotopes to these waiting point nuclei are also included in the
analysis. The nuclear structure part of the problem is described within a
deformed Skyrme Hartree-Fock + BCS + QRPA approach, which reproduces not
only the beta-decay half-lives but also the available Gamow-Teller strength
distributions, measured under terrestrial conditions. The various
sensitivities of the decay rates to both density and temperature are
discussed. In particular, we study the impact of contributions coming from 
thermally populated excited states in the parent nucleus, as well as
the competition between beta decays and continuum electron captures. 

\end{abstract}


\maketitle

\section{Introduction}

An accurate understanding of most astrophysical processes requires 
necessarily information from nuclear physics, which provides the
input to deal with network calculations and astrophysical
simulations (see \cite{langanke03,apra} and references therein).
Obviously, nuclear physics uncertainties will finally affect the
reliability of the description of those astrophysical processes.
This is especially relevant in the case of explosive phenomena,
which involve knowledge of the properties of exotic nuclei, not
well explored yet. Thus, most of the astrophysical simulations
of these violent events must be built on nuclear-model predictions
of limited quality and accuracy. This is in particular the case
of the X-ray bursts (XRBs) \cite{wallace,thielemann,schatz,woosley},
which are generated by a thermonuclear runaway in the hydrogen-rich
environment of an accreting neutron star that is fed from a red
giant binary companion close enough to allow for mass transfer.

Type I XRBs are typically characterized by a rapid increase in
luminosity generating burst energies of $10^{39}-10^{40}$ ergs, which
are typically a factor 100 larger than the steady luminosity.
The luminosity suffers a sharp raise of about $1-10$ s followed by a
gradual softening with time scales between 10 and 100 s. These bursts
are recurrent with time scales ranging from hours to days. The
properties of XRBs are particularly dependent on the accretion rate.
Typical accretion rates for type  I XRBs are about
$ 10^{-8}-10^{-9} M_{\odot}$ yr$^{-1}$.  Lower accretion rates lead to
weaker flashes while larger accretion rates lead to stable burning
on the surface of the neutron star.

The ignition of XRBs takes place when the temperature ($T$) and the
density ($\rho$) in the accreted disk become high enough to allow a
breakout from the hot CNO cycle. Peak conditions of $T=1-3$ GK and 
$\rho=10^6-10^7$ g cm$^{-3}$ are reached and eventually, this scenario
allows the development of the nucleosynthesis rapid proton capture
($rp$) process \cite{schatz,woosley,wormer,pruet}, which is
characterized by proton capture reaction rates that are orders of
magnitude faster than any other competing process, in particular
$\beta$-decay. It produces rapid nucleosynthesis on the proton-rich
side of stability toward heavier proton-rich nuclei reaching nuclei
with $A\gtrsim 100$, as it have been studied in Ref. \cite{schatz01},
where it was shown that the $rp$ process ends in a closed SnSbTe
cycle. It also explains the energy and luminosity profiles observed
in XRBs. 

Nuclear reaction network calculations, which may involve as much as
several thousand nuclear processes, are performed to follow the time 
evolution of the isotopic abundances, to determine the amount of
energy released by nuclear reactions, and to find the reaction path
for the $rp$ process 
\cite{wallace,thielemann,schatz,woosley,wormer,pruet,schatz01,ffn}. 
In general, the reaction path follows a series of fast proton-capture
reactions until the dripline is reached and further proton capture is
inhibited by a strong reverse photodisintegration reaction. At this
point, the process may only proceed through a beta decay or a less
probable double proton capture. Then the reaction flow has to wait for
a relatively slow $\beta$-decay and the respective nucleus is called
a waiting point (WP). The short time scale of the $rp$ process (around
100 s) makes highly significant any mechanism that may affect the
process in some seconds and the half-lives of the WP nuclei are of this
order. Therefore, the half-lives of the WP nuclei along the reaction path
determine the time scale of the nucleosynthesis process and the produced
isotopic abundances. In this respect, the weak decay rates of
neutron-deficient medium-mass nuclei under stellar conditions play a
relevant role to understand the $rp$ process.

Although the products of the nucleosynthesis $rp$ process are not
expected to be ejected from type I XRBs due to the strength of the
neutron star gravitational field, there are other speculative sites
for the occurrence of $rp$ processes. This is the case of core collapse
supernovae that might supply suitable physical conditions for the $rp$
process provided neutrino-induced reactions are included in the
nucleosynthesis calculations \cite{wanajo}. These reactions have to be
included to bypass the slow beta decays at the WP nuclei via capture
reactions of neutrons, which are created from the antielectron neutrino
absorption by free protons \cite{frohlich}. Contrary to the XRBs, these
scenarios will finally lead to the ejection of the nucleosynthetic
products and thus contribute to the galactic chemical evolution.

Since the pioneering work of Fuller, Fowler and Newman \cite{ffn}, 
where the general formalism to calculate weak-interaction rates in
stellar environments as a function of density and temperature was
introduced, improvements have been focused on the description of the
nuclear structure aspect of the problem. Different approaches to
describe the nuclear structure involved in the stellar weak decay
rates can be found in the literature. They are basically divided
into Shell Model  \cite{langanke00,langanke01} or quasiparticle
random phase approximation (QRPA) \cite{nabi,paar09,sarri_plb}
categories. Certainly, the nuclear structure problem involved in the
calculation of these rates must be treated in a reliable way. In
particular, this implies that the nuclear models should be able to
describe at least the experimental information available on the
decay properties (Gamow-Teller strength distributions and
$\beta$-decay half-lives) measured under terrestrial conditions.
Although these decay properties may be different at the high
$\rho$ and $T$ existing in $rp$ process scenarios, success in
describing the decay properties in terrestrial conditions is a
requirement for a reliable calculation of the weak decay rates in
more general conditions. With this aim in mind, we study here the
dependence of the decay rates on both  $\rho$ and $T$ using a QRPA
approach based on a selfconsistent deformed Hartree-Fock (HF) mean
field. Deformation has to be taken into account because the reaction
path in the $rp$ process crosses a region of highly deformed nuclei
around $A=70-80$. This nuclear model has been tested successfully
(see \cite{sarri_prc} and references therein) and reproduces very
reasonably the experimental information available on both bulk and
decay properties of medium-mass nuclei. In this work we focus our
attention to the even-even WP Ni, Zn, Ge, Se, Kr, Sr, Zr, Mo, Ru, 
Pd, Cd, and Sn isotopes, as well as to their closer even-even
neighbors.

The paper is organized as follows. In Section \ref{wdr} 
the weak decay rates are introduced as functions of density
and temperature and their nuclear structure and phase space
components are studied. Section \ref{results} contains the results.
First, we study the decay properties under terrestrial conditions,
and secondly as functions of both densities and temperatures at
the $rp$ process. Section \ref{conclusions} contains
the conclusions of this work.

\section{ Weak decay rates}  
\label{wdr}

There are several distinctions between terrestrial and stellar decay
rates caused by the effect of high  $\rho$ and $T$. The main effect of
$T$ is directly related to the thermal population of excited states in
the decaying nucleus, accompanied by the corresponding depopulation of
the ground states. The weak-decay rates of excited states can be
significantly different from those of the ground state and a
case by case consideration is needed. Another effect related to the
high $\rho$ and $T$ comes from the fact that atoms in these scenarios
are completely ionized and consequently electrons are no longer bound
to the nuclei, but forming a degenerate plasma obeying a Fermi-Dirac
distribution. This opens the possibility for continuum electron
capture ($cEC$), in contrast to the orbital electron capture ($oEC$)
produced by bound electrons in the atom under terrestrial conditions.
These effects make weak interaction rates in the stellar interior
sensitive functions of T and $\rho$, with $T=1.5$ GK and $\rho=10^6$ 
g cm$^{-3}$, as the most significant conditions for the $rp$ 
process \cite{schatz}.

The decay rate of the parent nucleus is given by

\begin{equation}
\lambda = \sum_i \lambda_i\, \frac{2J_i+1}{G} e^{-E_i/(kT)} \, ,
\label{population}
\end{equation}
where $G=\sum_i \left( 2J_i+1 \right) e^{-E_i/(kT)}$ is the partition
function, $J_i(E_i)$ is the angular momentum (excitation energy) of 
the parent nucleus state $i$, and thermal equilibrium is assumed. 
In principle, the sum extends over all populated states in the parent
nucleus up to the proton separation energy. However, since the range
of temperatures for the $rp$ process peaks at $T=1.5$ GK ($kT\sim 300$ 
keV), only a few low-lying excited states are expected to contribute
significantly in the decay. Specifically, we consider in this work
all the collective low-lying excited states below 1 MeV \cite{ensdf}.
Two-quasiparticle excitations in even-even nuclei will appear at
an excitation energy above 2 MeV, which is a typical energy to break
a pair in these isotopes. Hence, they can be safely neglected at
these temperatures. 
As an example, the maximum population appears for the lowest of these
states ($E_{2^+}=261$ keV in $^{76}$Sr), which at $T$=1.5 GK is $12\%$,
while the ground state still contributes with $88\%$. 

The decay rate for the parent state $i$ is given by

\begin{equation}
\lambda _i = \sum_f \lambda_{if}\, ,
\end{equation}
where the sum extends over all the states in the final nucleus
reached in the decay process. The rate $\lambda_{if}$ from the initial
state $i$ to the final state $f$ is given by

\begin{equation}
\lambda _{if} = \frac{\ln 2}{D}  B_{if}\Phi_{if} (\rho,T)\, ,
\end{equation}
where $D=6146$ s. This expression is decomposed into a nuclear
structure part $B_{if}$ that contains the transition probabilities
for allowed Fermi (F) and Gamow-Teller (GT) transitions,

\begin{equation}
B_{if}=B_{if}(GT)+ B_{if}(F)\, ,
\end{equation}
and a phase space factor $\Phi_{if}$, which is a sensitive function
of $\rho$ and $T$.
The theoretical description of both $B_{if}$ and $\Phi_{if}$
are explained in the next subsections.

\subsection{Nuclear Structure}

The nuclear structure part of the problem is described within the
QRPA formalism. Various approaches have been developed in the past
to describe the spin-isospin nuclear excitations in QRPA 
\cite{krum,hamamoto,moller,hir,homma,borzov,paar04,fracasso,petro,
sarri_74,sarri_pp,sarri_odd}.
In this subsection we show briefly the theoretical framework used in
this paper to describe the nuclear part of the decay rates in the
neutron-deficient nuclei considered in this work. More details of the
formalism can be found in Refs. \cite{sarri_74,sarri_pp,sarri_odd}.

The method starts with a self-consistent deformed Hartree-Fock mean 
field formalism obtained with Skyrme interactions, including
pairing correlations. The single-particle energies, wave functions,
and occupation probabilities are generated from this mean field.
In this work we have chosen the Skyrme force SLy4 \cite{sly4} as a
representative of the Skyrme forces. This particular force includes
some selected properties of unstable nuclei in the adjusting procedure
of the parameters. It is one of the most successful Skyrme forces
and has been extensively studied in the last years.

The solution of the HF equation is found by using the formalism 
developed in Ref. \cite{vautherin}, assuming time reversal and axial 
symmetry. The single-particle wave functions are expanded in terms 
of the eigenstates of an axially symmetric harmonic oscillator in 
cylindrical coordinates, using twelve major shells. The method also 
includes pairing between like nucleons in BCS approximation with 
fixed gap parameters for protons and neutrons, which are determined
phenomenologically from the odd-even mass differences involving
the experimental binding energies \cite{audi}. 

The potential energy curves are analyzed as a function of
the quadrupole deformation $\beta$, 

\begin{equation}
\beta = \sqrt{\frac{\pi}{5}}\frac{Q_0}{A\langle r^2 \rangle}\, ,
\label{beta_quadru}
\end{equation}
written in terms of the mass quadrupole moment $Q_0$ and the mean 
square radius $\langle r^2 \rangle$. For that purpose, constrained 
HF calculations are performed with a quadratic constraint 
\cite{constraint}. The HF energy is minimized under the constraint 
of keeping fixed the nuclear deformation. Calculations for GT 
strengths are performed subsequently for the various equilibrium
shapes of each nucleus, that is, for the solutions, in general
deformed, for which minima are obtained in the energy curves. 
Since decays connecting different shapes are disfavored, similar
shapes are assumed for the ground state of the parent nucleus and
for all populated states in the daughter nucleus.
The validity of this assumption was discussed for example in 
Refs. \cite{krum,homma}. 

To describe GT transitions, a spin-isospin residual interaction is
added to the mean field and treated in a deformed proton-neutron QRPA.
This interaction contains two parts, a particle-hole ($ph$) and a
particle-particle ($pp$). The interaction in the $ph$ channel is 
responsible for the position and structure of the GT resonance 
\cite{homma,sarri_gese} and it can be derived consistently from 
the same Skyrme interaction used to generate the mean field, through 
the second derivatives of the energy density functional with respect 
to the one-body densities. The $ph$ residual interaction is finally 
expressed in a separable form by averaging the resulting contact 
force over the nuclear volume \cite{sarri_74}. 
The $pp$ part is a neutron-proton pairing force in the $J^\pi=1^+$ 
coupling channel, which is also introduced as a separable force 
\cite{hir,sarri_pp}.
The strength of the $pp$ residual interaction in this theoretical
approach is not derived self-consistently from the SLy4 force 
used to obtain the mean field, but nevertheless it has 
been fixed in accordance to it. This strength is usually fitted to 
reproduce globally the experimental half-lives. Various 
attempts have been done in the past to fix this strength \cite{homma},
arriving to expressions that depend on the model used to describe
the mean field, Nilsson model in the above reference. 
In previous works 
\cite{sarri_pp,sarri_gese,sarri_pb,sarri_wp,sarri_pere}
we have studied the sensitivity of the GT strength distributions to 
the various ingredients contributing to the deformed QRPA-like 
calculations, namely to the nucleon-nucleon effective force, to 
pairing correlations, and to residual interactions. We found different 
sensitivities to them. In this work, all of these ingredients have been 
fixed to the most reasonable choices found previously \cite{sarri_prc}
and mentioned above. In particular we use the coupling strengths 
$\chi ^{ph}_{GT}=0.15$ MeV and $\kappa ^{pp}_{GT} = 0.03$ MeV.

The proton-neutron QRPA phonon operator for GT excitations in
even-even nuclei is written as

\begin{equation}
\Gamma _{\omega _{K}}^{+}=\sum_{\pi\nu}\left[ X_{\pi\nu}^{\omega _{K}}
\alpha _{\nu}^{+}\alpha _{\bar{\pi}}^{+}+Y_{\pi\nu}^{\omega _{K}}
\alpha _{\bar{\nu}} \alpha _{\pi}\right]\, ,  \label{phon}
\end{equation}
where $\alpha ^{+}\left( \alpha \right) $ are quasiparticle creation
(annihilation) operators, $\omega _{K}$ are the QRPA excitation 
energies, and $X_{\pi\nu}^{\omega _{K}},Y_{\pi\nu}^{\omega _{K}}$ the 
forward and backward amplitudes, respectively. For even-even nuclei 
the allowed GT transition amplitudes in the intrinsic frame
connecting the QRPA ground state
$\left| 0\right\rangle \ \ \left( \Gamma _{\omega _{K}} \left| 0
\right\rangle =0 \right)$ to one-phonon states $\left| \omega _K
\right\rangle \ \ \left( \Gamma ^+ _{\omega _{K}} \left| 0
\right\rangle = \left| \omega _K \right\rangle \right)$,
are given by

\begin{equation}
\left\langle \omega _K | \sigma _K t^{\pm} | 0 \right\rangle =
\mp M^{\omega _K}_\pm \, ,\quad K=0,1\, ,
\label{intrinsic}
\end{equation}
where
\begin{eqnarray}
M_{-}^{\omega _{K}}&=&\sum_{\pi\nu}\left( q_{\pi\nu}X_{\pi
\nu}^{\omega _{K}}+ \tilde{q}_{\pi\nu}Y_{\pi\nu}^{\omega _{K}} \right) , \\
M_{+}^{\omega _{K}}&=&\sum_{\pi\nu}\left( \tilde{q}_{\pi\nu} 
X_{\pi\nu}^{\omega _{K}}+ q_{\pi\nu}Y_{\pi\nu}^{\omega _{K}}\right) \, ,
\end{eqnarray}
with
\begin{equation}
\tilde{q}_{\pi\nu}=u_{\nu}v_{\pi}\Sigma _{K}^{\nu\pi },\ \ \
q_{\pi\nu}=v_{\nu}u_{\pi}\Sigma _{K}^{\nu\pi},
\label{qs}
\end{equation}
$v'$s are occupation amplitudes ($u^2=1-v^2$) and
$\Sigma _{K}^{\nu\pi}$ spin matrix elements connecting neutron
and proton states with spin operators
\begin{equation}
\Sigma _{K}^{\nu\pi}=\left\langle \nu\left| \sigma _{K}\right|
\pi\right\rangle \, .
\end{equation}

The GT strength for a transition from an initial state $i$ to a
final state $f$ is given by

\begin{equation}
B_{if}(GT^{\pm} )= \frac{1}{2J_i+1} \left( \frac{g_A}{g_V} 
\right)_{\rm eff} ^2 \langle f || \sum_j^A \sigma_j t^{\pm}_j 
|| i \rangle ^2 \, ,
\end{equation}
where $(g_A/g_V)_{\rm eff} = 0.74 (g_A/g_V)_{\rm bare}$  is an
effective quenched value. For the  transition 
$I_iK_i (0^+0) \rightarrow I_fK_f (1^+K)$ in the laboratory
system, the energy distribution of the GT strength  
$B_{\omega}(GT^\pm )$ is expressed in terms of the intrinsic amplitudes
in Eq. (\ref{intrinsic}) as

\begin{eqnarray}
B_{\omega}(GT^\pm )& =&  \left( \frac{g_A}{g_V} \right)_{\rm eff} ^2 
\sum_{\omega_{K}} \left[ \left\langle \omega_{K} \left| \sigma_0t^\pm 
\right| 0 \right\rangle ^2 \delta_{K,0} \right.  \nonumber  \\
&& \left. +  2 \left\langle \omega_{K} \left| \sigma_1t^\pm \right|
0 \right\rangle ^2 \delta_{K,1}  \right] \, .
\label{bgt}
\end{eqnarray}
To obtain this expression, the initial and final states in the
laboratory frame have been expressed in terms of the intrinsic states
using the Bohr-Mottelson factorization \cite{bm}.

Concerning Fermi transitions, the Fermi operator is the isospin ladder
operator $T_{\pm}$, which commutes with the nuclear part of the Hamiltonian
excluding the small Coulomb component. Then, superallowed Fermi transitions 
($0^+ \rightarrow 0^+$) only occur between members of an isospin multiplet.
The Fermi strength is narrowly concentrated in the isobaric analog state
(IAS) of the ground state of the decaying nucleus.
Thus, neglecting effects from isospin mixing one has

\begin{equation}
B_{if}(F^{\pm}) = \frac{1}{2J_i+1}\langle f || \sum_j^A t^{\pm}_j || i 
\rangle ^2  = T(T+1)-T_{z_i}T_{z_f} \, ,
\end{equation}
where $T$ is the nuclear isospin and $T_{z}=(N-Z)/2$ its third component.
The $B_{if}(F^+)$ strength of our concern here reduces to $B(F^+)=(Z-N)=2$
for the $(T,T_z)=(1,-1)$ isotopes in the decay $(Z,N)\rightarrow (Z-1,N+1)$
with $Z=N+2$. For these transitions the excitation energy of the IAS in
the daughter nucleus is given by \cite{pruet,ffn}

\begin{equation}
E_{IAS}=(ME)_i-(ME)_f + 0.7824 - \Delta E_C \, {\rm MeV},
\end{equation}
where $ME$ is the atomic mass excess. The Coulomb displacement energy 
$\Delta E_C$  between pairs of isobaric analog levels is given by
\begin{equation}
\Delta E_C = 1.4144 {\bar Z}/A^{1/3} -0.9127 {\rm MeV}\, ,
\end{equation}
where $\bar{Z}=(Z_i+Z_f)/2$. This expression was obtained in 
Ref. \cite{antony} from a fitting to data corresponding to levels
with isospin $T=1$. In any case, Fermi transitions are only important
for the $\beta^+$ decay of neutron-deficient light nuclei with 
$Z> N$ ($T_z<0$), where the IAS can be reached energetically.
Thus, although they have been considered in the calculations of the
terrestrial half-lives, only the dominant GT transitions are included
in the stellar decay rates.

\subsection{Phase Space Factors}

The phase space factor contains two components, electron capture ($EC$)
and $\beta^+$ decay

\begin{equation}
\Phi_{if}=\Phi^{EC}_{if}+\Phi^{\beta^+}_{if}\, .
\end{equation}

In the case of $\beta^+/EC$ decay in the laboratory, $EC$ arises from
orbital electrons in the atom and the phase space factor is given
by \cite{gove}

\begin{equation}
\Phi^{oEC}=\frac{\pi}{2} \sum_x q_x^2 g_x^2B_x\, ,
\end{equation}
where $x$ denotes the atomic subshell from which the electron is captured,
$q$ is the neutrino energy, $g$ is the radial component of the bound-state
electron wave function at the nucleus, and $B$ stands for other exchange
and overlap corrections \cite{gove}.

In $rp$-process stellar scenarios, the phase space factor for $cEC$ is
given by

\begin{eqnarray}
\Phi^{cEC}_{if}&=&\int_{\omega_\ell}^{\infty} \omega p (Q_{if}+\omega)^2
F(Z,\omega) \nonumber \\
&& \times S_{e^-}(\omega) \left[ 1-S_{\nu}(Q_{if}+\omega)\right] d\omega \, .
\label{phiec}
\end{eqnarray}

The phase space factor for positron emission $\beta^+$ process is
given by 

\begin{eqnarray}
\Phi^{\beta^+}_{if}&=&\int _{1}^{Q_{if}} \omega p  
(Q_{if}-\omega)^2 F(-Z+1,\omega) \nonumber \\
&& \times \left[ 1-S_{e^+}(\omega)\right]
\left[ 1-S_{\nu}(Q_{if}-\omega)\right] d\omega \, .
\label{phib}
\end{eqnarray}
In these expressions $\omega$ is the total energy of the positron in 
$m_ec^2$ units, $p=\sqrt{\omega ^2 -1}$ is the momentum in $m_e c$ units,
and $Q_{if}$ is the total energy available in $m_e c^2$ units

\begin{equation}
Q_{if}=\frac{1}{m_ec^2}\left( M_p-M_d+E_i-E_f \right) \, ,
\end{equation}
which is is written in terms of the nuclear masses of parent ($M_p$)
and daughter ($M_d$) nuclei and their excitation energies $E_i$ and
$E_f$, respectively. $F(Z,\omega)$ is the Fermi function \cite{gove}
that takes into account the distortion of the $\beta$-particle wave
function due to the Coulomb interaction.

\begin{equation}
F(Z,\omega ) = 2(1+\gamma) (2pR)^{-2(1-\gamma)} e^{\pi y}
\frac{|\Gamma (\gamma+iy)|^2}{[\Gamma (2\gamma+1)]^2}\, ,
\end{equation}
where $\gamma=\sqrt{1-(\alpha Z)^2}$ ; $y=\alpha Z\omega /p$ ; 
$\alpha$ is the fine structure constant and $R$ the nuclear radius. 
The lower integration limit in the $cEC$ expression is given by
${\omega_\ell}=1$ if $Q_{if}> -1$, or ${\omega_\ell}=|Q_{if}|$ if 
$Q_{if}< -1$.

$S_{e^-}$, $S_{e^+}$, and $S_\nu$, are the electron, positron, and neutrino
distribution functions, respectively. Its presence inhibits or enhances
the phase space available. In $rp$ scenarios the commonly accepted
assumptions \cite{schatz} state that $S_\nu=0$ since neutrinos and
antineutrinos can escape freely from the interior of the star and then
they do not block the emission of these particles
in the capture or decay processes. Positron distributions become only
important at higher $T$ ($kT > 1$ MeV) when positrons appear via 
pair creation, but at the temperatures considered here we take $S_{e^+}=0$.
The electron distribution is described as a Fermi-Dirac distribution

\begin{equation}
S_{e}=\frac{1}{\exp \left[ \left(\omega -\mu_e\right)/(kT)\right] +1} \, ,
\end{equation}
assuming that nuclei at these temperatures are fully ionized and 
electrons are not bound to nuclei.
The chemical potential $\mu_e$ is determined from the expression
\begin{equation}
\rho Y_e = \frac{1}{\pi^2 N_A}\left( \frac{m_e c}{\hbar}\right) ^3 
\int_0^{\infty} (S_e - S_{e^+}) p² dp \, ,
\end{equation}
in (mol/cm$^3$) units. $\rho$ is the baryon density (g/cm$^3$),
$Y_e$ is the electron-to-baryon ratio (mol/g), and $N_A$ is Avogadro's
number (mol$^{-1}$).

Under the assumptions $S_{e^+}=S_\nu =0$ mentioned above, the phase space
factors for $\beta^+$ decay in Eq. (\ref{phib}) are independent of
the density and temperature. The only dependence of the $\beta^+$
decay rates on $T$ arises from the thermal population of excited
parent states. On the other hand, the phase space factor for $cEC$
in Eq.  (\ref{phiec}) is a function of both $\rho Y_e$ and $T$, through
the electron distribution $S_{e^-}$. 

The phase space factors increase with $Q_{if}$ and thus the decay rates
are more sensitive to the strength $B_{if}$ located at low excitation
energies of the daughter nucleus. It is also interesting to notice the
relative importance of both  $\beta^+$ decay and electron capture phase
space factors (see Fig. 3 in Ref.\cite{sarri_plb}). 
In general, the former dominates at sufficiently high $Q_{if}$
(low excitation energies in the daughter nucleus), while
the latter is always dominant at sufficiently low $Q_{if}$
(high excitation energies in the daughter nucleus).

The $\beta$-decay half-life in the laboratory is obtained by summing
all the allowed transition strengths to states in the daughter nucleus
with excitation energies lying below the corresponding $Q_{EC}$ energy,
and weighted with the phase space factors,

\begin{equation}
T_{1/2}^{-1}=\frac{\lambda}{\ln 2}=\frac{1}{D} \sum_{0 < E_f < Q_{EC}} 
\left[ B_{if}(GT)+B_{if}(F) \right]  \Phi_{if}^{\beta^+/oEC} \, ,
\label{t12}
\end{equation}
where the $Q_{EC}$ energy is given by

\begin{equation}
Q_{EC} = M_p-M_d+m_e = Q_{\beta^+} + 2m_e\, .
\end{equation}

\section{Results for weak decay rates}
\label{results}

In this section we present first the results for the potential energy
curves. Then, we show the results for the decay properties, GT strength
distributions and $\beta$-decay half-lives, under terrestrial conditions
comparing them with the available experimental information.
Finally, we present the results for the stellar weak decay rates under
density and temperature conditions implied in the $rp$ process.

\subsection{Potential Energy Curves}

\begin{figure*}
\centering
\includegraphics[width=0.8\textwidth]{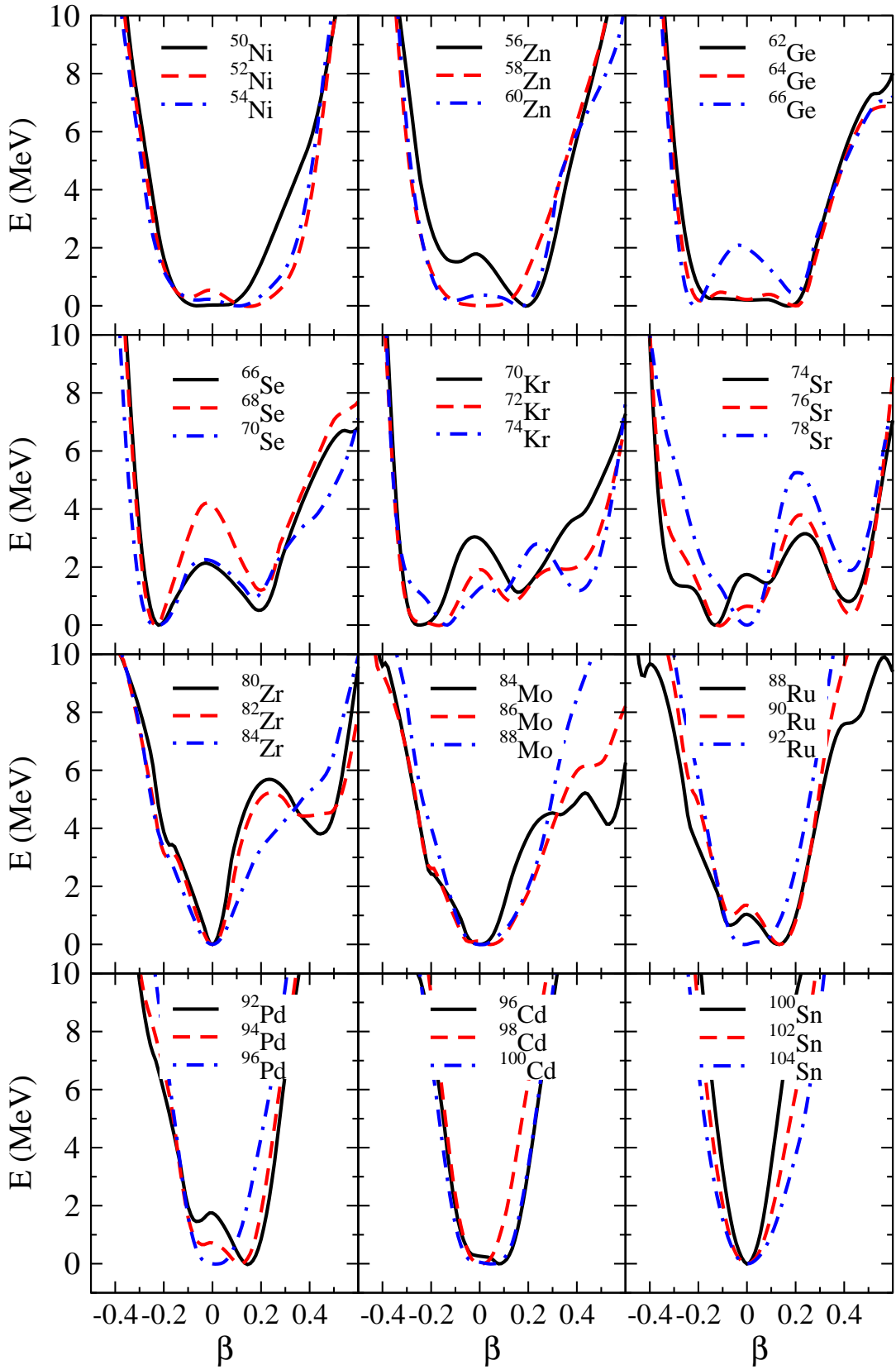}
\caption{(Color online) Potential energy curves for the even-even isotopes
considered in this work obtained from constrained HF+BCS calculations with
the Skyrme force SLy4.}
\label{fig_eq}
\end{figure*}

In Fig. \ref{fig_eq} we can see the potential energy curves for the
even-even Ni, Zn, Ge, Se, Kr, Sr, Zr, Mo, Ru, Pd, Cd, and Sn nuclei in
the vicinity of the $N=Z$ isotopes considered in this work. We show the
energies relative to that of the ground state plotted as a function of
the quadrupole deformation $\beta$ in Eq. (\ref{beta_quadru}). They
are obtained from constrained HF+BCS calculations with the Skyrme
force SLy4 \cite{sly4}. 

The nuclei studied here cover a whole proton shell ranging from magic
number $Z=28$ (Ni isotopes) up to magic number $Z=50$ (Sn isotopes).
The isotopes considered are the predicted WP nuclei, which in most
cases correspond to $N=Z$, and their neighbor isotopes.
 
Then, it is expected that the lighter and heavier nuclei close to
$Z=28$ and $Z=50$, respectively, have a tendency to be spherical.
The spherical shapes in these isotopes show sharply peaked profile
that become shallow minima as one moves away from $Z=28$ or $Z=50$,
and finally deformed shapes are developed as one approaches mid-shell
nuclei. The profiles of the latter exhibit a rich structure giving
raise to shape coexistence when various minima at close energies are
located at different deformations.

It is also worth mentioning the correlations observed between mirror
nuclei interchanging the number of neutrons and protons.
Thus, we see the remarkable similarity between the profiles of 
$^{66}$Ge $(Z=32,N=34)$ and $^{66}$Se $(Z=34,N=32)$, between
$^{70}$Se $(Z=34,N=36)$ and $^{70}$Kr $(Z=36,N=34)$, and between
$^{74}$Kr $(Z=36,N=38)$ and $^{74}$Sr $(Z=38,N=36)$.

These results are in qualitative agreement with similar ones obtained
in this mass region from different theoretical approaches. Just
to give some examples, shape transition and shape coexistence were
discussed in $A\sim 80$ nuclei within a configuration-dependent
shell-correction approach based on a deformed Woods-Saxon potential
\cite{naza85}. Relativistic mean field calculations in this mass
region have also been reported in Ref. \cite{relat}. Nonrelativistic
calculations are also available from both Skyrme 
\cite{nonrel_skyrme_1,nonrel_skyrme_2,nonrel_skyrme_3} and 
Gogny \cite{nonrel_gogny} forces, as well as from the complex
VAMPIR approach \cite{petrovici}.

Experimental evidence of shape coexistence in this mass region
has become available in the last years 
\cite{wood,piercey,chandler,becker,fisher,bouchez,gade,gorgen05,
clement,davies,andreoiu,singh,hurst,gorgen07,ljungvall,obertelli},
and by now this is a well established characteristic feature in the
neutron-deficient $A=70-80$ mass region.

\subsection{Laboratory Gamow-Teller strength and half-lives}

While the half-lives give only a limited information of the decay
(different strength distributions may lead to the same half-life), the
strength distribution contains all the information. It is of great
interest to study the decay rates under stellar $rp$ conditions using
a nuclear structure model that reproduces the strength distributions
and half-lives under terrestrial conditions.

In the next figures, we show the results obtained for the energy
distributions of the GT strength corresponding to the equilibrium
shapes for which we obtained minima in the potential energy curves
in Fig. \ref{fig_eq}. The GT strength is plotted versus the excitation
energy of the daughter nucleus $E_{ex}=E_f$ (MeV).

\begin{figure*}
\includegraphics[width=0.50\textwidth]{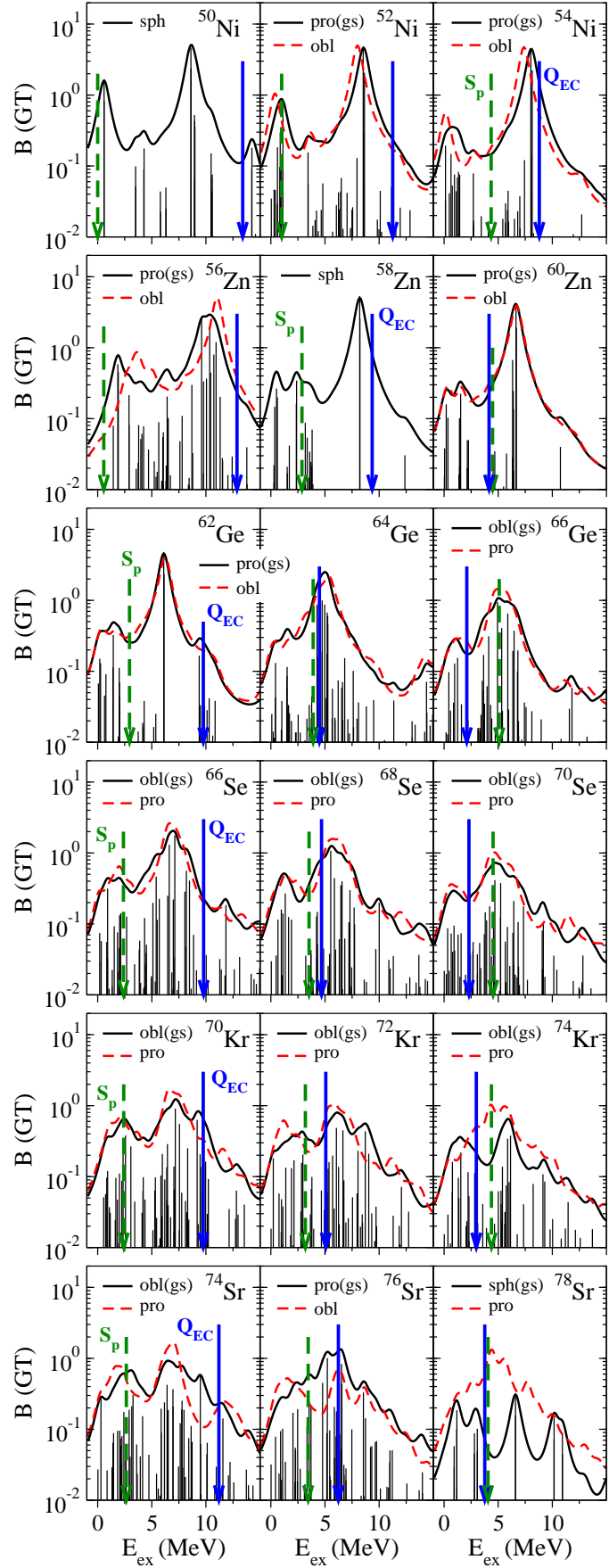}
\caption{(Color online) Calculated GT strength distributions for Ni, Zn, Ge,
Se, Kr, and Sr isotopes obtained from their ground states, as well as from the
shape coexisting states. The individual strengths correspond to the ground
states, whereas folded distributions are shown for the various configurations
considered in each isotope. $Q_{EC}$ values and proton separation energies $S_p$
are shown by vertical lines.}
\label{fig_bgt1}
\end{figure*}

\begin{figure*}
\includegraphics[width=0.50\textwidth]{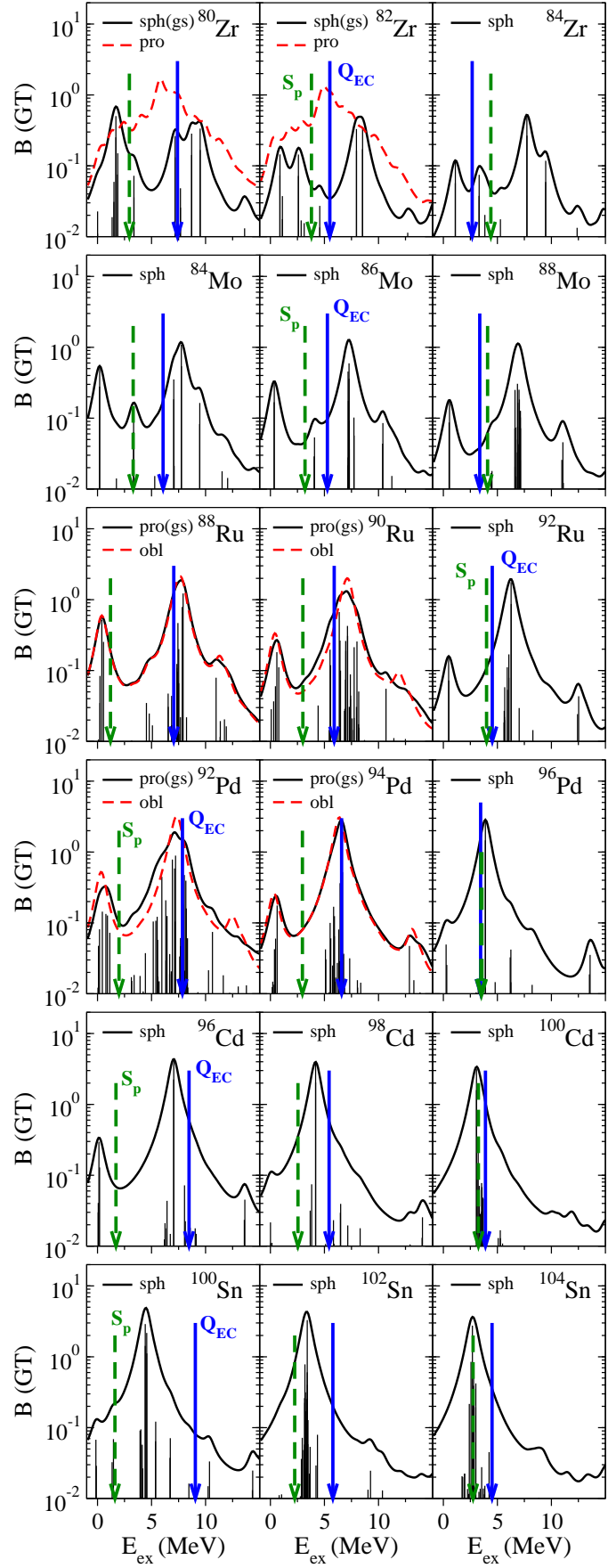}
\caption{(Color online) Same as in Fig. \ref{fig_bgt1}, 
but for Zr, Mo, Ru, Pd, Cd, and Sn isotopes.}
\label{fig_bgt2}
\end{figure*}

Fig. \ref{fig_bgt1} (\ref{fig_bgt2}) contains the results for the
isotopes Ni, Zn, Ge, Se, Kr, and Sr (Zr, Mo, Ru, Pd, Cd, and Sn). We
show the energy distributions of the individual GT strengths in the
case of the ground state shapes. We also show the continuous
distributions for both  ground state and possible shape isomers,
obtained by folding the strength with 1 MeV width Breit-Wigner
functions. The vertical arrows show the $Q_{EC}$ energy,
as well as the proton separation energy in the daughter nucleus,
both taken from experiment \cite{audi}.

It is worth noticing that in general both deformations produce quite
similar GT strength distributions on a global scale. 
The main exceptions correspond to the comparison between spherical
and deformed shapes, where clear differences can be observed.
In any case, the small differences among the various shapes 
at the low energy tails (below the $Q_{EC}$) of the GT strength
distributions lead to sizable effects in the $\beta$-decay half-lives. 
These differences can be better seen because of the logarithmic scale.

Experimental information on GT strength distributions are mainly
available for $^{72}$Kr \cite{piqueras}, $^{74}$Kr \cite{poirier},
$^{76}$Sr \cite{nacher}, and $^{102,104}$Sn \cite{karny} isotopes,
where $\beta^+$-decay experiments have been performed with total
absorption spectroscopy techniques, allowing the extraction of the
GT strength in practically the whole $Q$-energy window. In Ref. 
\cite{sarri_prc} a comparison between similar calculations to those
in this work and the experimental data for Kr and Sr isotopes was
carried out. In general, good agreement with experiment was found
and this was one of the reasons to extrapolate this type of
calculations to stellar environments, as well as to other WP nuclei.

\begin{figure*}
\includegraphics[width=0.9\textwidth]{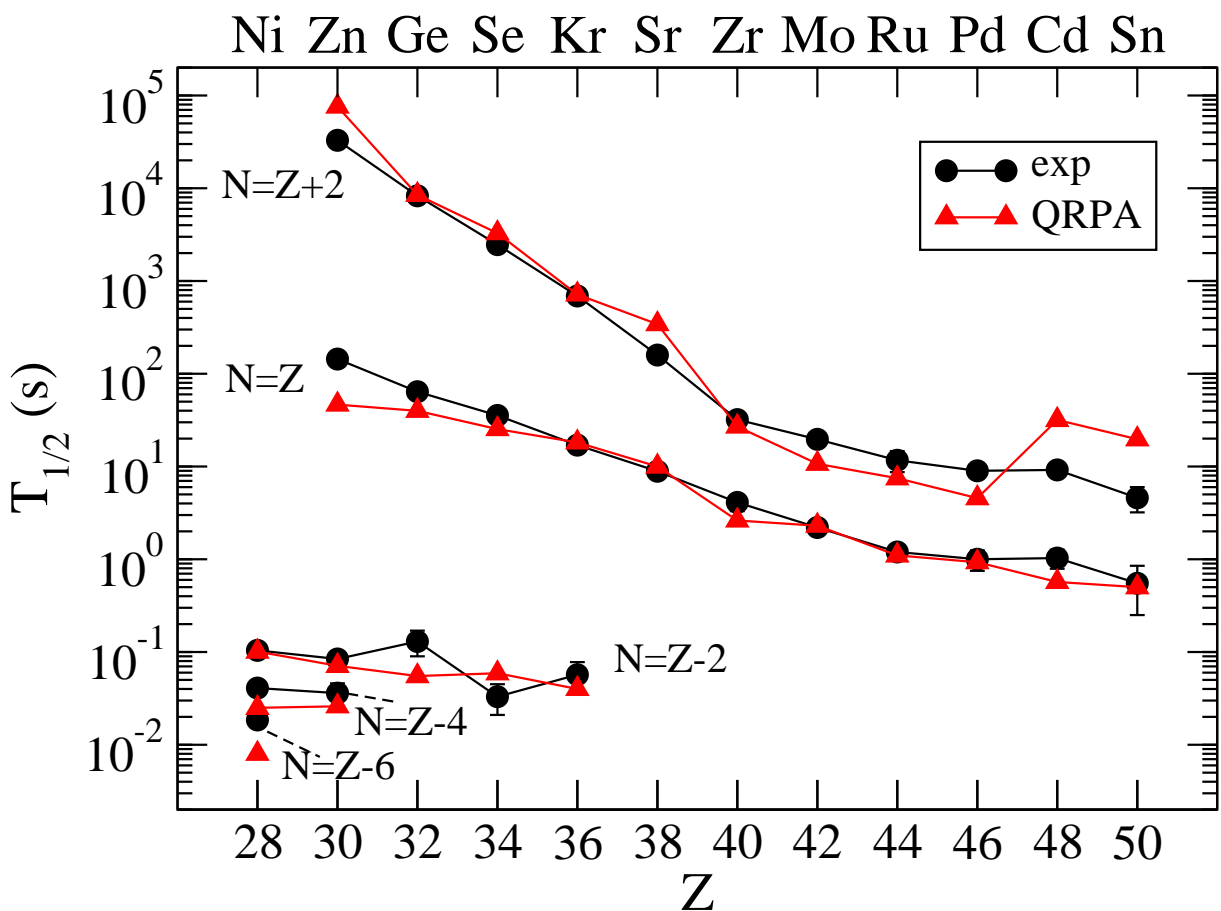}
\caption{ (Color online) Calculated QRPA half-lives compared to experimental
values.}
\label{fig_t12}
\end{figure*}

Measurements of the decay properties (mainly half-lives) of nuclei
in this mass region have been reported in the last years
\cite{karny,oinonen,kienle,faestermann,wohr,kankainen,kavatsyuk,dossat,
bazin,stoker,weber,elomaa}. 
The calculation of the half-lives in Eq.(\ref{t12}) involves the knowledge
of the GT strength distribution and of the $Q_{EC}$ values. In this work
experimental values for $Q_{EC}$ are used. 
They are taken from Ref. \cite{audi} or from the Jyv\"askyl\"a mass
database \cite{weber,jyvaskyla},  when available.
In Fig. \ref{fig_t12} the measured half-lives are compared to the QRPA
results obtained from the equilibrium deformations of the various isotopes.
In general good agreement for the $N=Z$ WP is obtained. Also for the more
stable $N=Z+2$ the agreement is very reasonable, except for the heavier
Cd an Sn isotopes, where the half-lives are overestimated. The half-lives
of the more exotic isotopes are fairly well described by QRPA.

\subsection{Stellar weak decay rates}

\begin{figure*}
\includegraphics[width=0.8\textwidth]{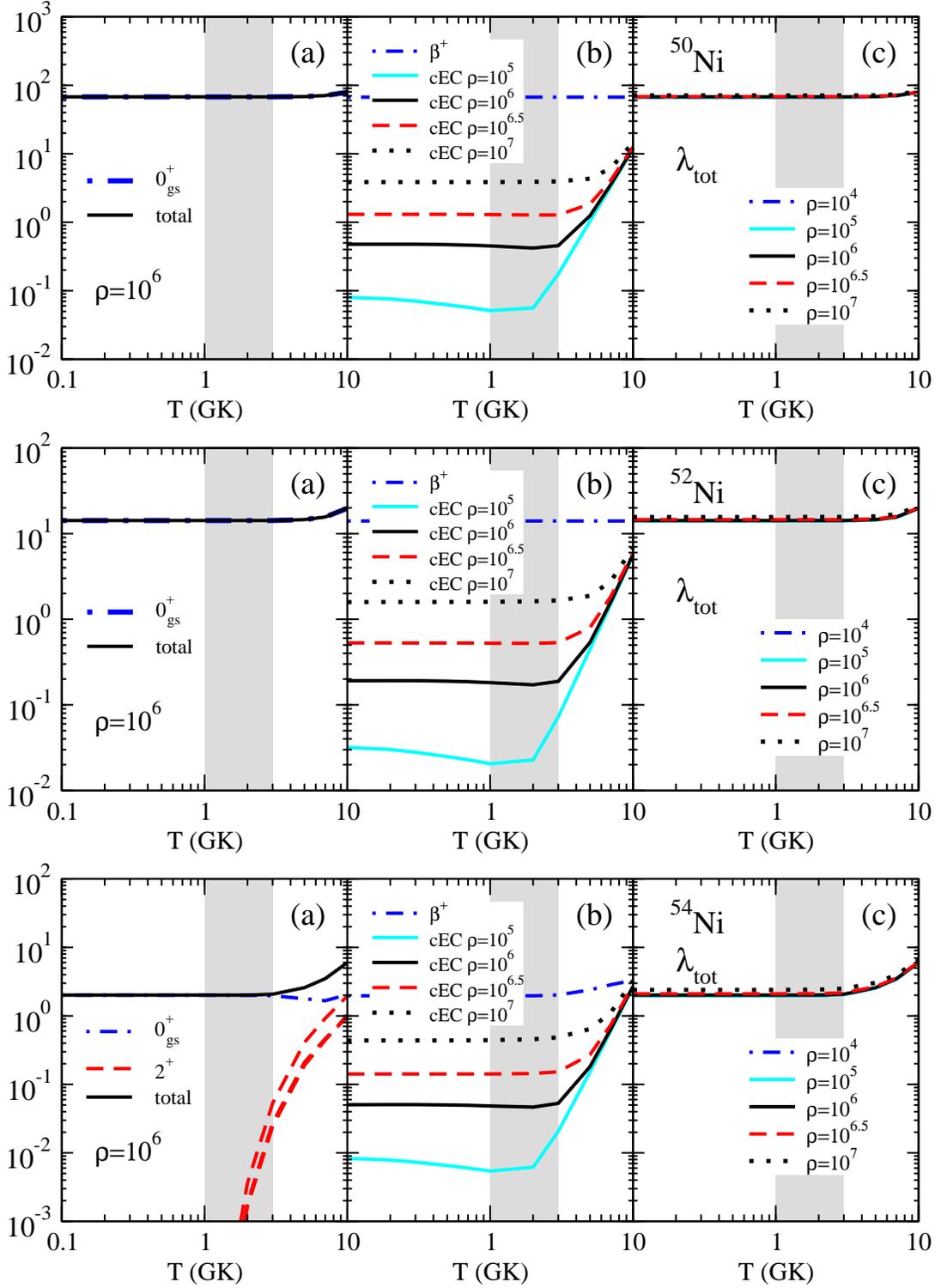}
\caption{(Color online) Decay rates ($s^{-1}$)of $^{50,52,54}$Ni isotopes as a function
of the temperature $T$ (GK). (a) Decomposition of the total rates into their
contributions from the decays of the ground and excited $2^+$ states. 
(b) Decomposition of the rates into their $\beta^+$ and  $cEC$  components
evaluated at different densities. (c) Total rates at various densities.
The label $\rho$ stands for $\rho Y_e$ (mol/cm$^3$) (see text).}
\label{fig_ni}
\end{figure*}

\begin{figure*}
\includegraphics[width=0.8\textwidth]{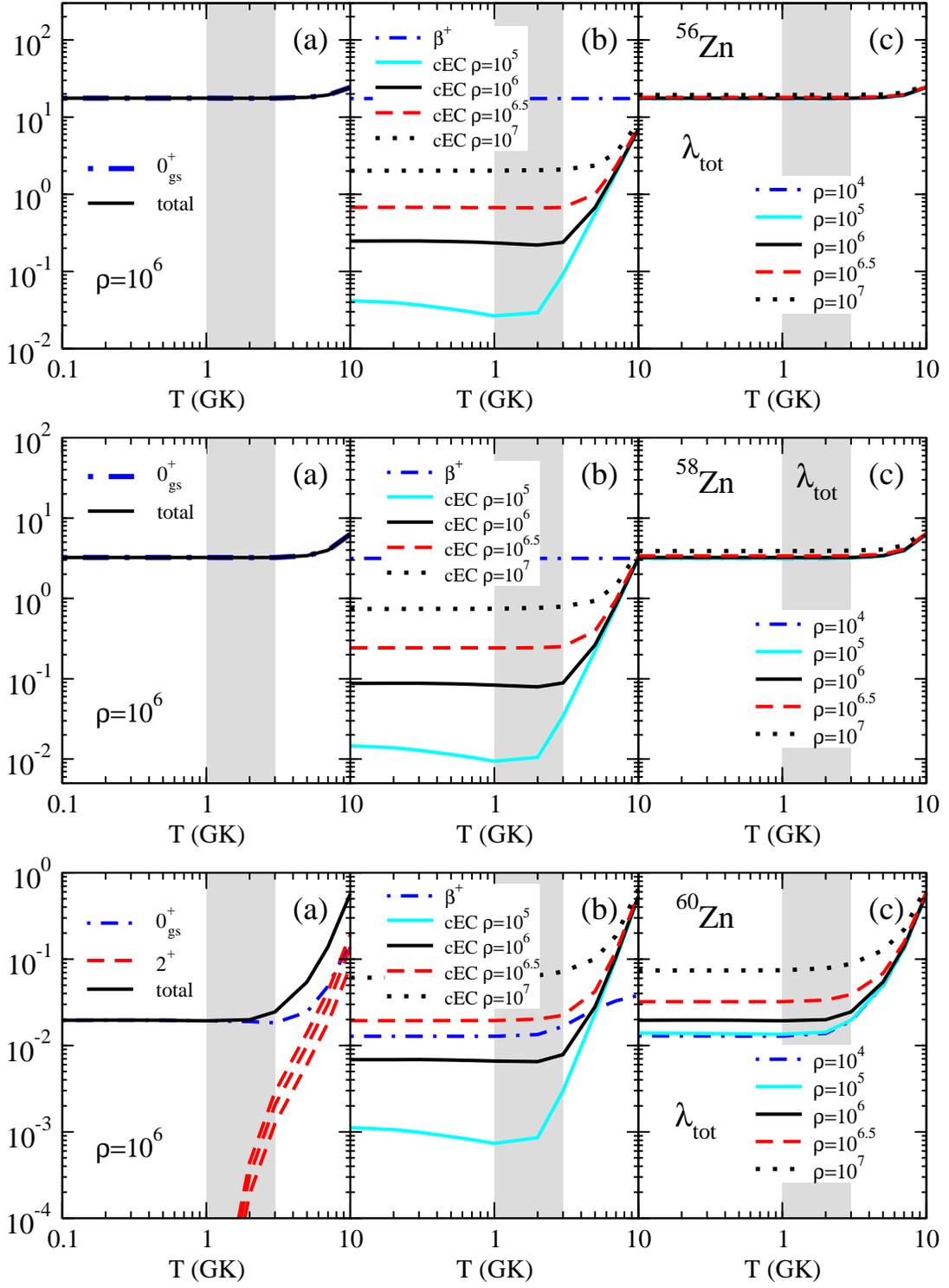}
\caption{(Color online) Same as in Fig. \ref{fig_ni}, but for  $^{56,58,60}$Zn
isotopes.}
\label{fig_zn}
\end{figure*}

\begin{figure*}
\includegraphics[width=0.8\textwidth]{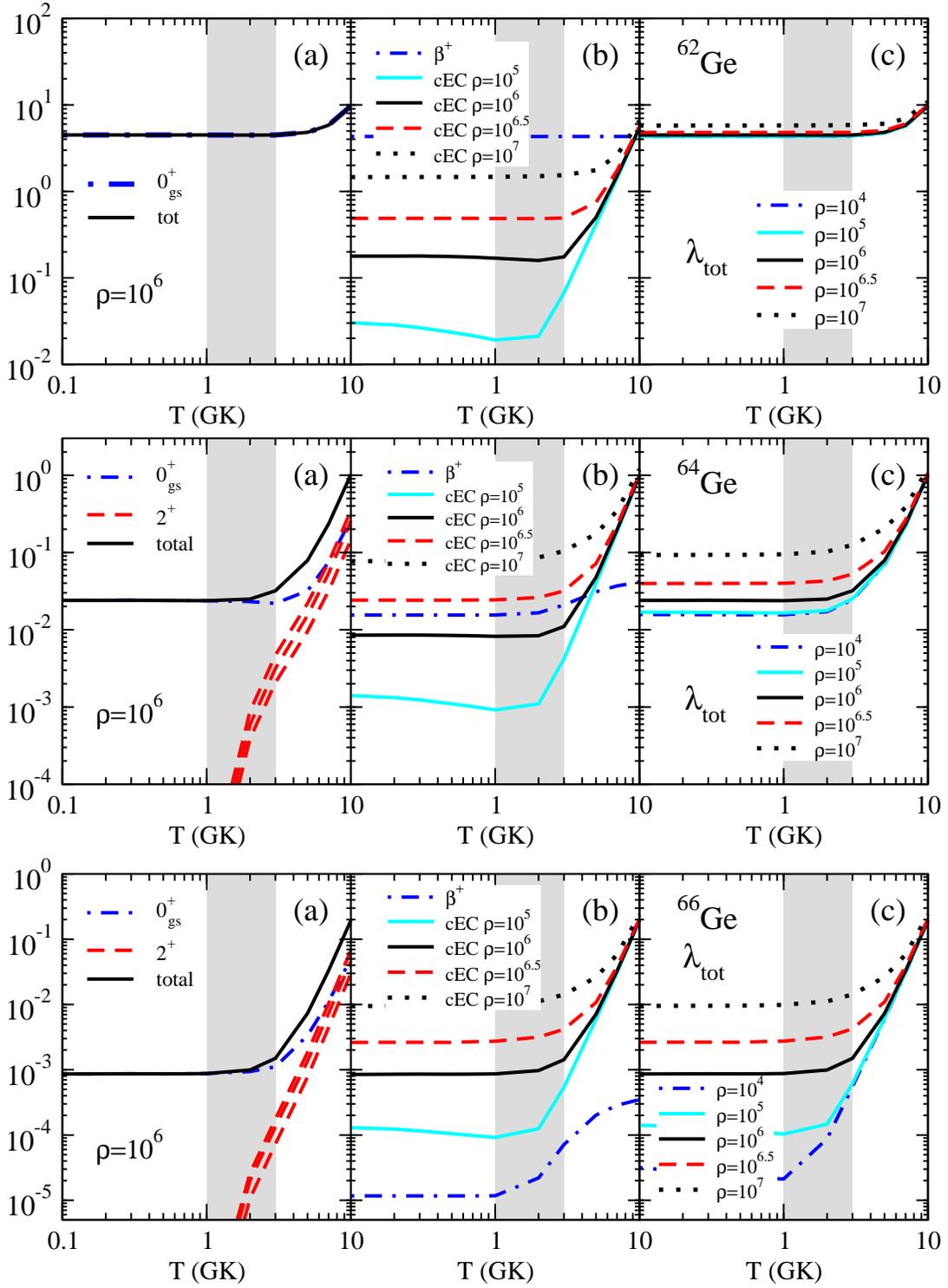}
\caption{(Color online) Same as in Fig. \ref{fig_ni}, but for  $^{62,64,66}$Ge
isotopes.}
\label{fig_ge}
\end{figure*}

\begin{figure*}
\includegraphics[width=0.8\textwidth]{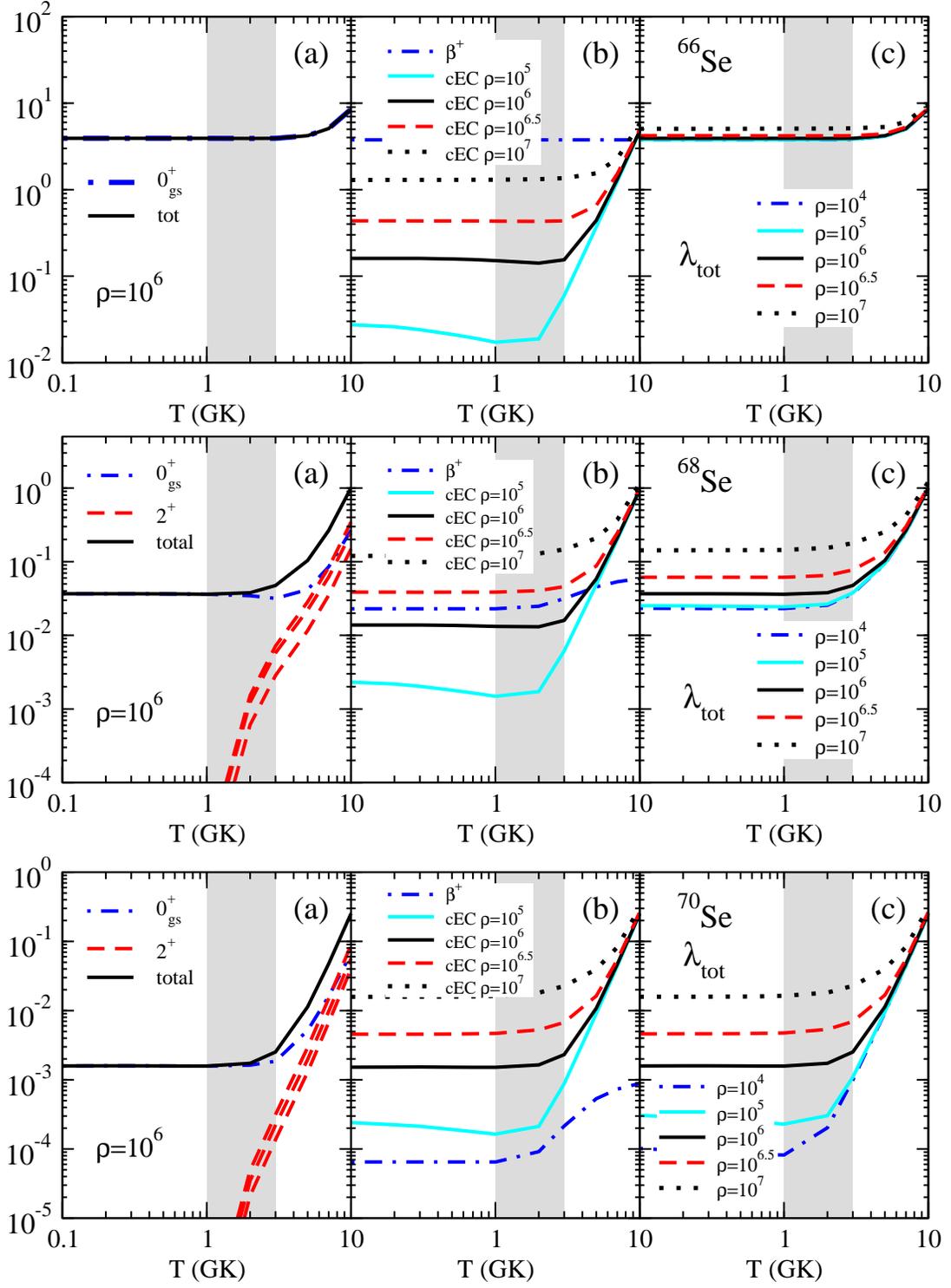}
\caption{(Color online) Same as in Fig. \ref{fig_ni}, but for  $^{66,68,70}$Se
isotopes.}
\label{fig_se}
\end{figure*}

\begin{figure*}
\includegraphics[width=0.8\textwidth]{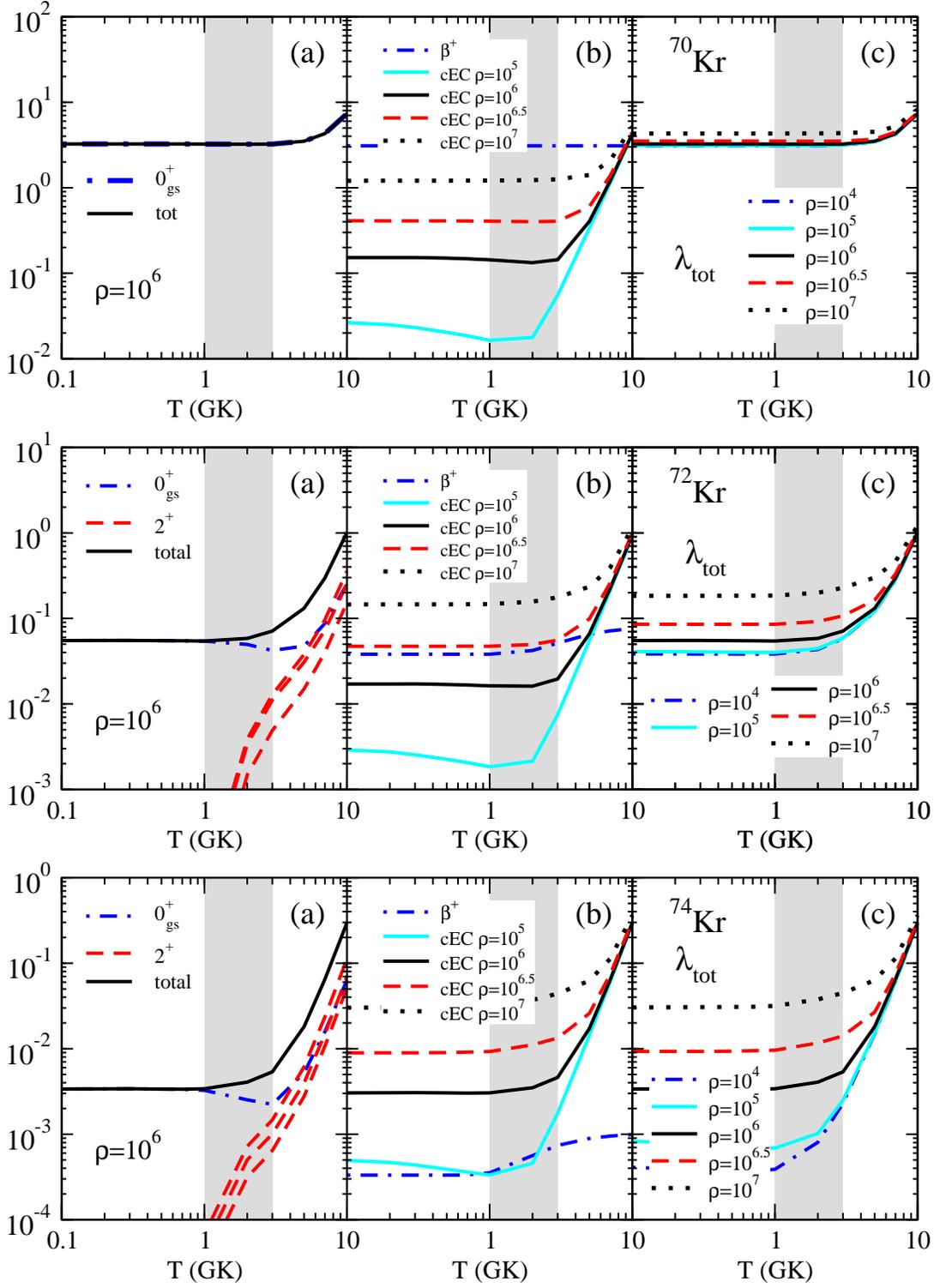}
\caption{(Color online) Same as in Fig. \ref{fig_ni}, but for  $^{70,72,74}$Kr
isotopes.}
\label{fig_kr}
\end{figure*}

\begin{figure*}
\includegraphics[width=0.8\textwidth]{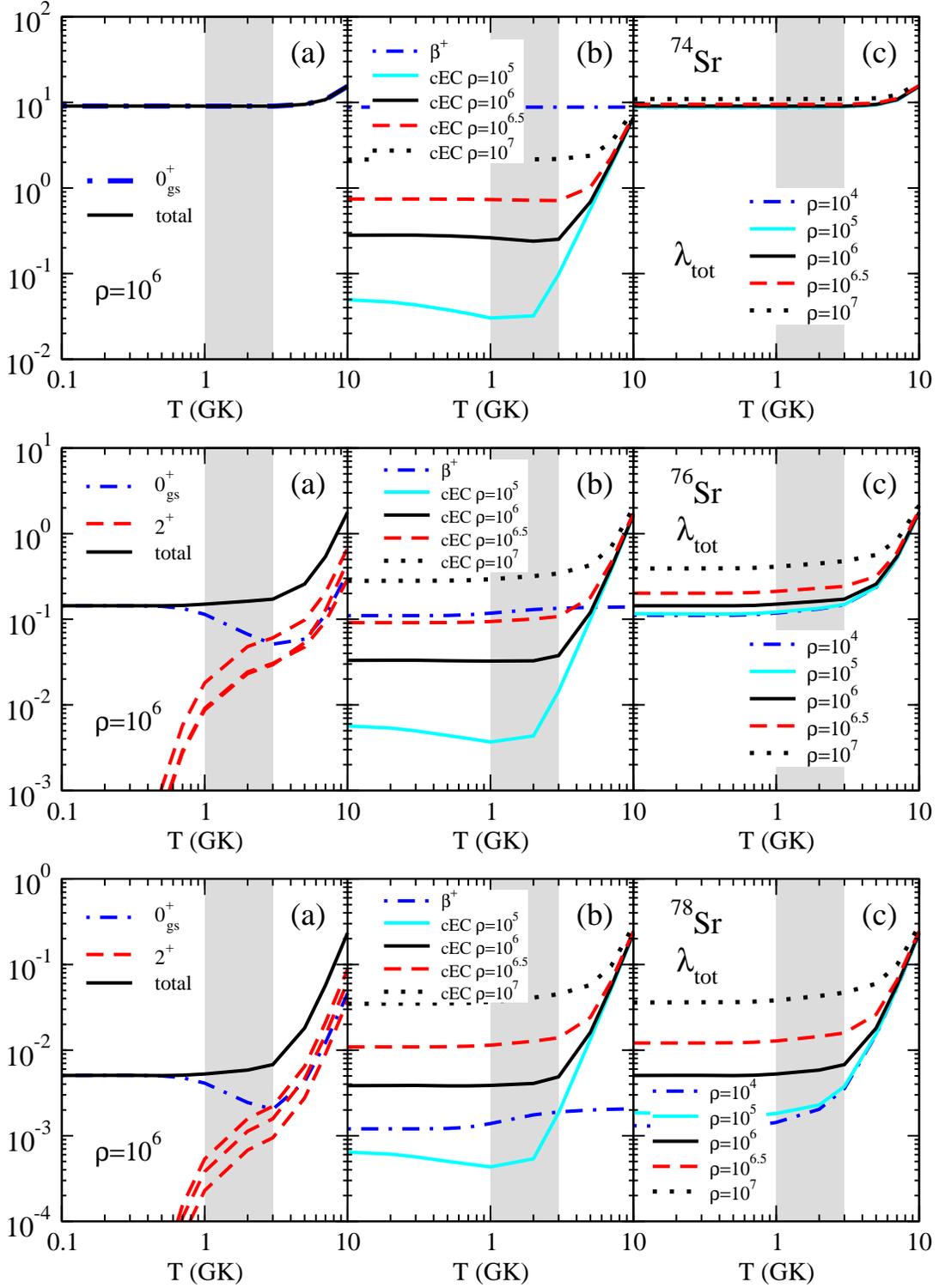}
\caption{(Color online) Same as in Fig. \ref{fig_ni}, but for  $^{74,76,78}$Sr
isotopes.}
\label{fig_sr}
\end{figure*}

\begin{figure*}
\includegraphics[width=0.8\textwidth]{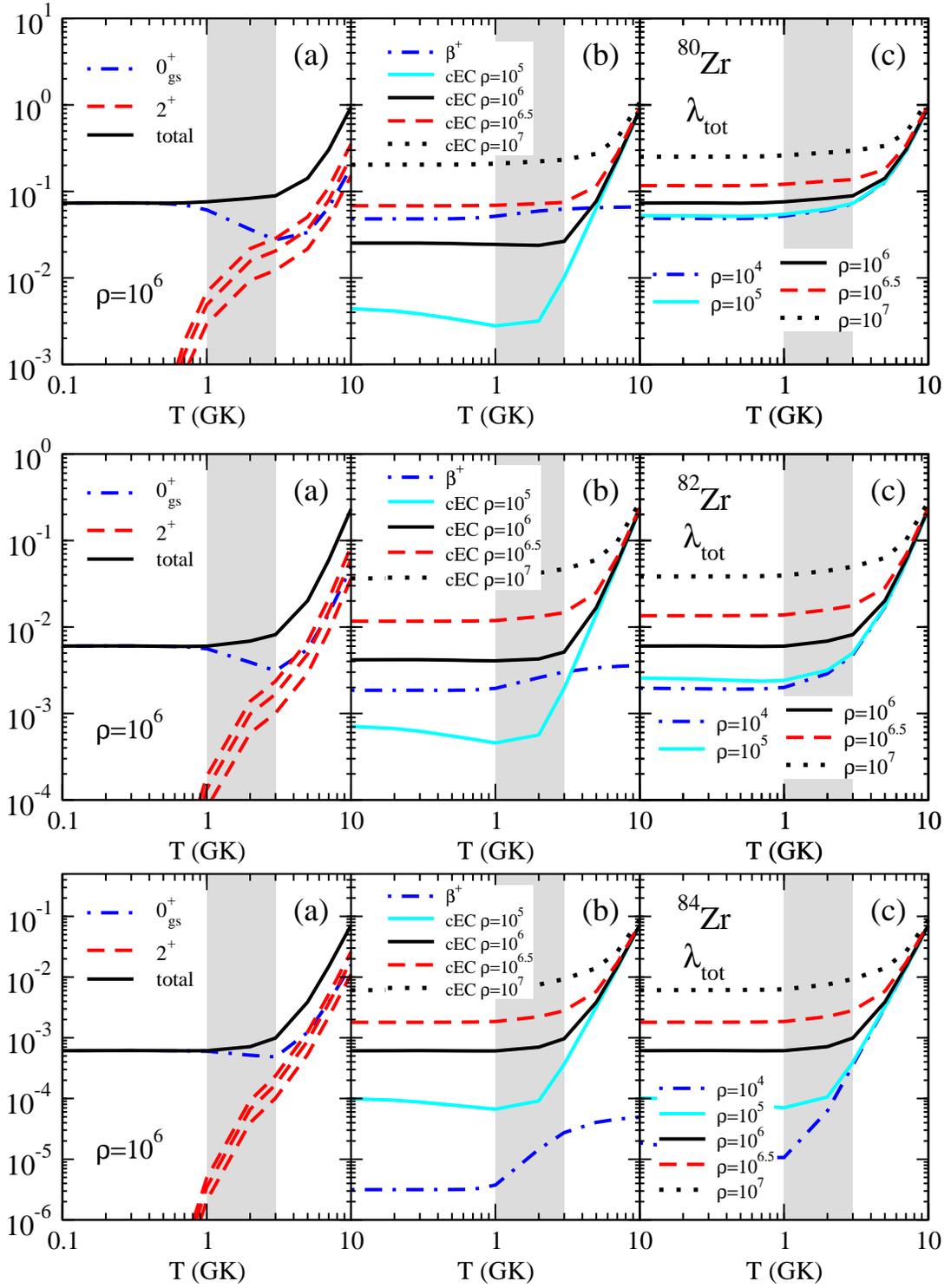}
\caption{(Color online) Same as in Fig. \ref{fig_ni}, but for  $^{80,82,84}$Zr
isotopes.}
\label{fig_zr}
\end{figure*}

\begin{figure*}
\includegraphics[width=0.8\textwidth]{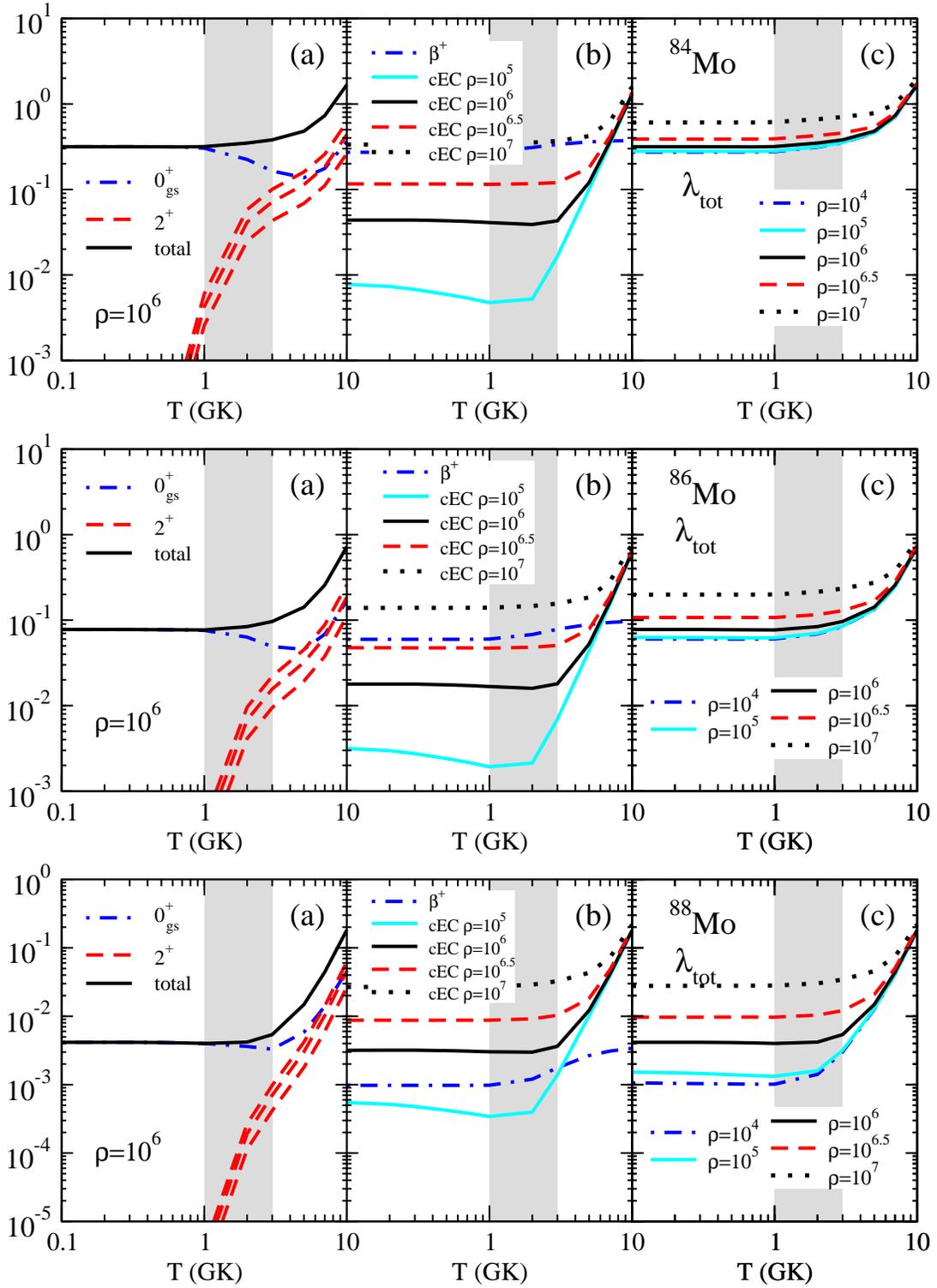}
\caption{(Color online) Same as in Fig. \ref{fig_ni}, but for  $^{84,86,88}$Mo
isotopes.}
\label{fig_mo}
\end{figure*}

\begin{figure*}
\includegraphics[width=0.8\textwidth]{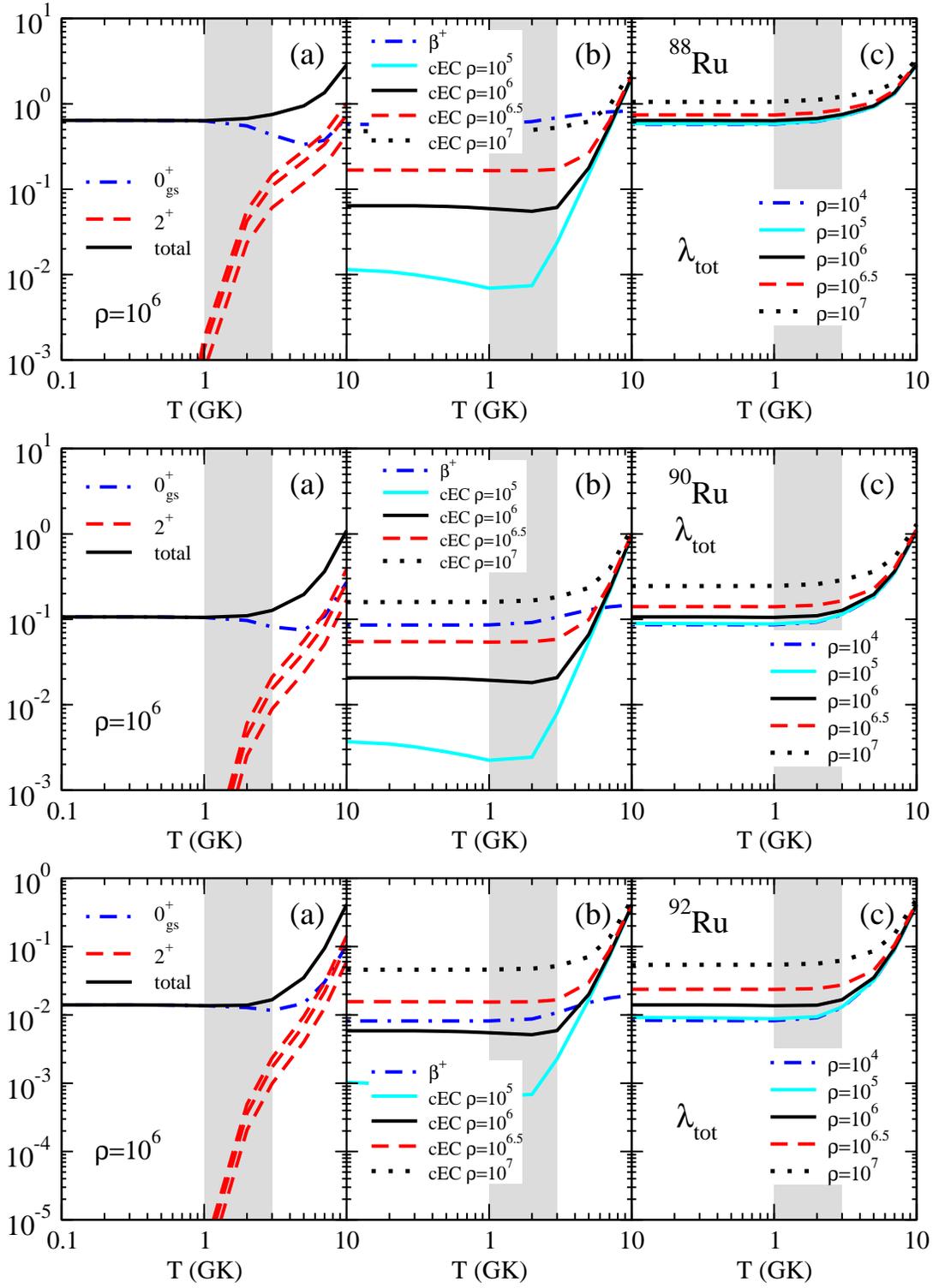}
\caption{(Color online) Same as in Fig. \ref{fig_ni}, but for  $^{88,90,92}$Ru
isotopes.}
\label{fig_ru}
\end{figure*}

\begin{figure*}
\includegraphics[width=0.8\textwidth]{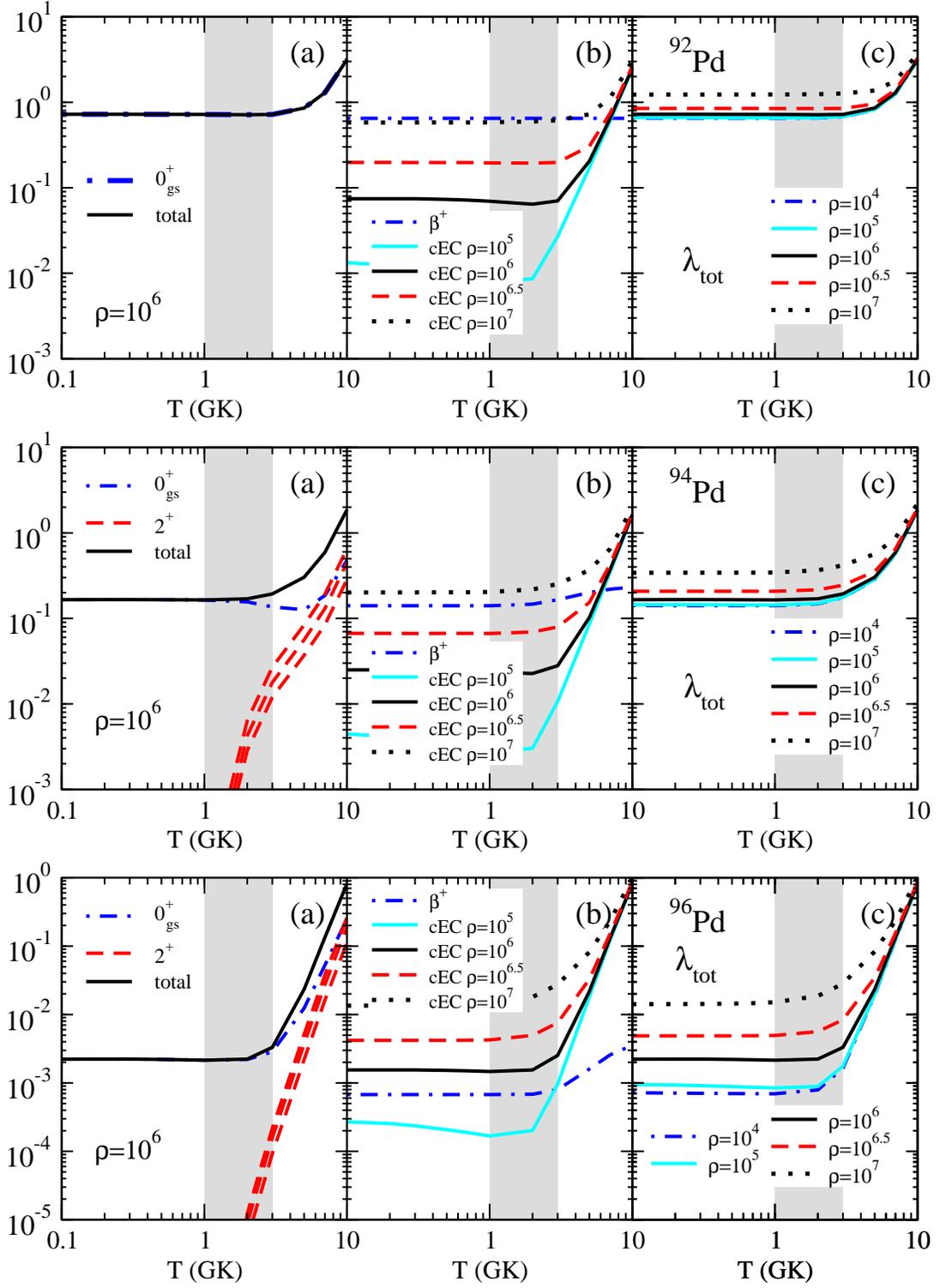}
\caption{(Color online) Same as in Fig. \ref{fig_ni}, but for  $^{92,94,96}$Pd
isotopes.}
\label{fig_pd}
\end{figure*}

\begin{figure*}
\includegraphics[width=0.8\textwidth]{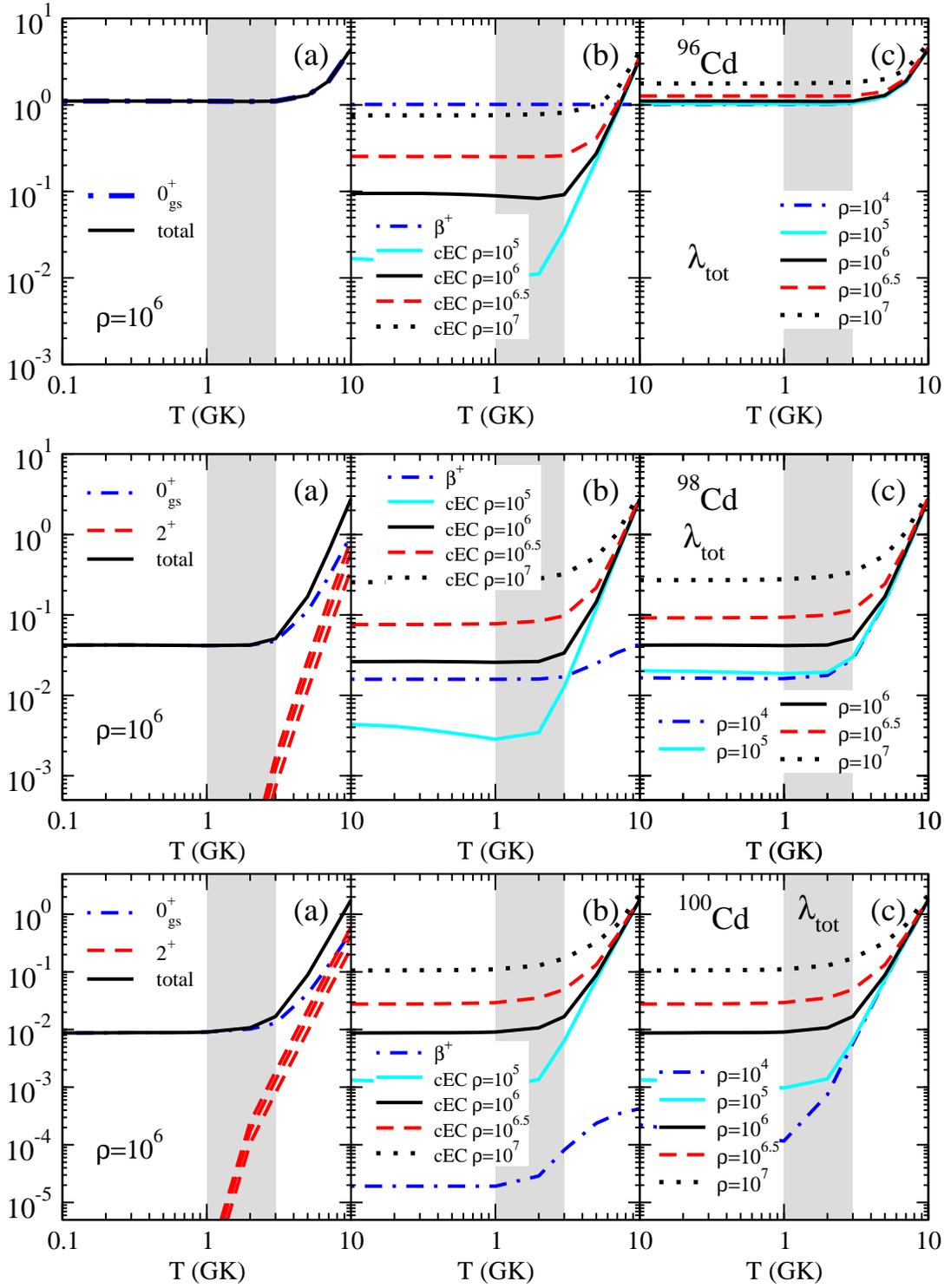}
\caption{(Color online) Same as in Fig. \ref{fig_ni}, but for  $^{96,98,100}$Cd
isotopes.}
\label{fig_cd}
\end{figure*}

\begin{figure*}
\includegraphics[width=0.8\textwidth]{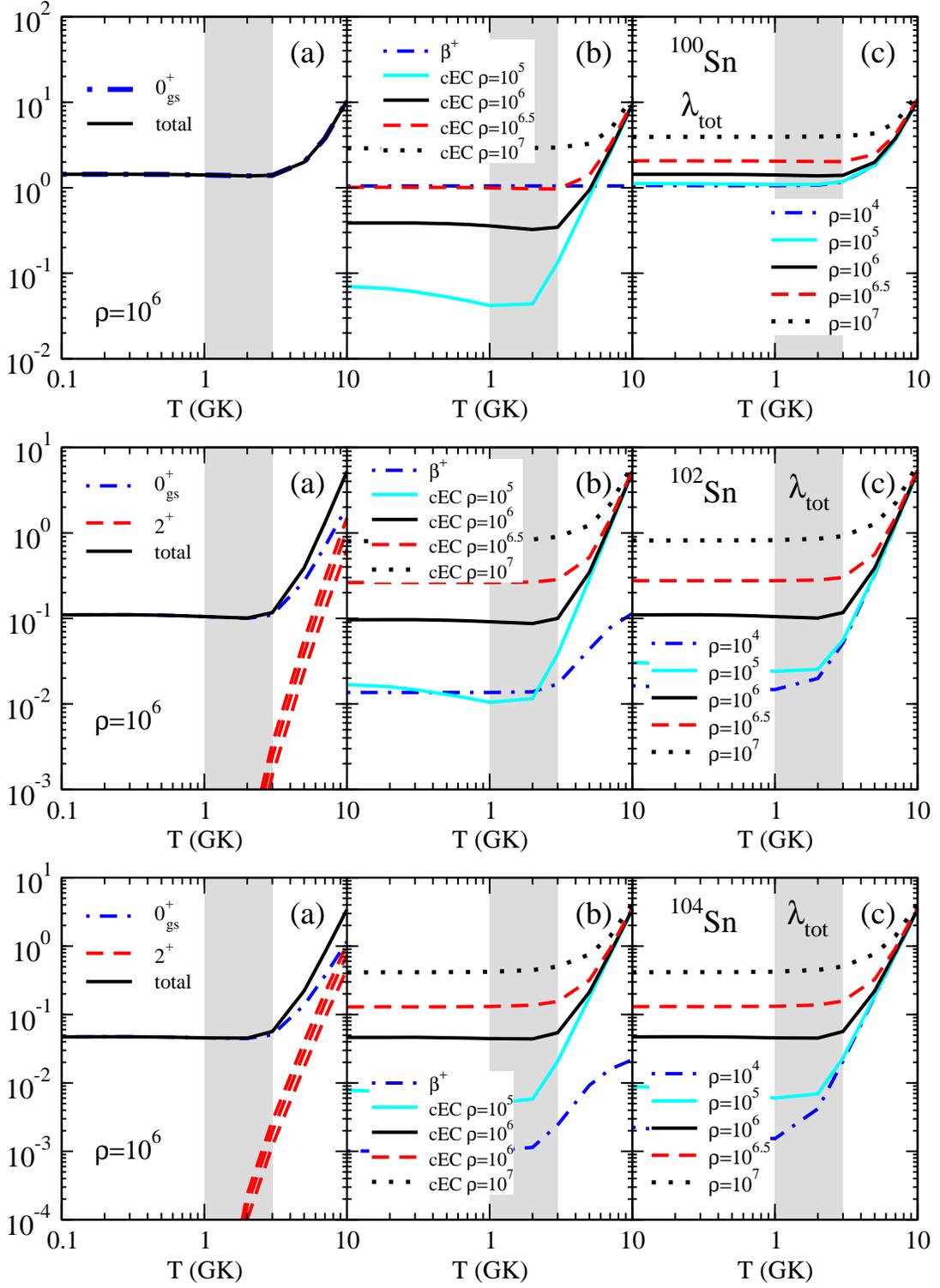}
\caption{(Color online) Same as in Fig. \ref{fig_ni}, but for  $^{100,102,104}$Sn
isotopes.}
\label{fig_sn}
\end{figure*}

Figures \ref{fig_ni}-\ref{fig_sn} show the decay rates as a function
of the temperature $T$. On the left-hand side (a) one can see the
decomposition of the total rates into their contributions from the
decay of the the ground state $0^+_{\rm gs}\rightarrow 1^+$ and from
the decay of the excited state $2^+ \rightarrow 1^+,2^+,3^+$ in the
parent nucleus. The middle panel (b) contains the decomposition of
the rates into their  $\beta^+$ and  $cEC$  components evaluated at
various densities ($\rho Y_e$). On the right-hand side (c) the total
rates for various densities are presented. The gray area is the relevant
range $T=1-3$ GK for the $rp$ process. Each figure contains the
results for three isotopes. The results corresponding to the more
exotic ones are displayed on top, whereas the results corresponding
to the more stable isotopes appear on the bottom. In the middle we
find the intermediate isotopes, which in most cases correspond to
the WP nuclei.

The results decomposed into their contributions from various parent
states (a) show that the decay from the ground state is always dominant
at the temperatures within the gray area of interest. The contributions
of the decays from excited states increase with $T$, as they become more
and more thermally populated, but in general they do not represent 
significant contributions to the total rates and can be neglected in
most cases. Nevertheless, there are a few cases where these contributions
should not be ignored, which correspond to those cases where the
excitation energy of the $2^+$ excited state is very low. This is the
case of the middle-shell nuclei Kr, Sr, Zr, and Mo, where the
contributions of the low-lying excited states compete with those
of the ground state already at temperatures in the range of $rp$
process. The effect on the rates of the decay from excited $0^+_2$ 
states was also considered in Ref. \cite{sarri_plb} in the case of
Kr and Sr isotopes. It was concluded that in general their relative
impact is again very small in the total rates at these temperatures.

Concerning the competition between  $\beta^+$ and $cEC$ rates (b) one
should distinguish between different isotopes. Thus, the more exotic
isotopes appearing on the top of the figures show a clear dominance of
the $\beta^+$ rates over the  $cEC$ ones that can be neglected except
at very high densities beyond $rp$-process conditions. On the other
hand, the opposite is true with respect to the more stable isotopes
on the bottom, where the  $\beta^+$ rates are completely negligible.
The origin of these features can be understood from the behavior of
the phase space factors as a function of the available energy $Q_{if}$.
As it was mentioned above and discussed in Ref. \cite{sarri_plb},
more exotic nuclei with larger $Q_{if}$ values favor  $\beta^+$ because
of the larger phase space factors, while the opposite is true for more
stable nuclei with smaller  $Q_{if}$ values.

The interesting cases occur in the middle panels that correspond in
most cases to the $N=Z$ WP nuclei. Here, there is a competition between
$\beta^+$ and $cEC$ rates that depends on the nucleus, on the
temperature, and on the density $\rho Y_e$. One can see that for large enough
densities, $cEC$ becomes dominant at any $T$. For low densities,
$\beta^+$ rates dominate at low $T$, while $cEC$ dominates at higher
$T$, but in general there is a competition that must be analyzed
case by case.

Finally, the total rates in (c) are a consequence of the competition
between $\beta^+$ and $cEC$ rates mentioned above. Since the $\beta^+$
decay rate is independent of the density and depends on $T$ only through
the contributions from excited parent states, the total rates are
practically constant for the more exotic isotopes in the upper figures,
only modulated by the small contribution from $cEC$.
In the central isotopes the rates are the result of the competition 
discussed in (b), and finally in the heavier isotopes (lower figures)  
we can see that the total rates are practically  due to $cEC$
with little contribution from  $\beta^+$.
Tables containing $\beta^+$, $cEC$, and total decay rates for all the
isotopes considered in this work are available in Ref. \cite{epaps}.

\section{Summary and Conclusions}
\label{conclusions}

In summary, the weak decay rates of waiting point and neighbor
nuclei from Ni up to Sn have been investigated at temperatures and
densities where the $rp$ process takes place. The nuclear structure
has been described within a microscopic QRPA approach based on a
selfconsistent Skyrme-Hartree-Fock-BCS mean field that includes
deformation. This approach reproduces both the experimental
half-lives and the more demanding GT strength distributions
measured under terrestrial conditions in this mass region. 

The relevant ingredients to describe the rates have been analyzed.
We have studied the contributions to the decay rates coming from
excited states in the parent nucleus which are populated as $T$
raises. It is found that they start to play a role above $T=1-2$
GK and that for isotopes with low-lying excited states, their
contributions can be comparable to those of the ground states.
Concerning the contributions from the continuum electron capture
rates, it is found that they are enhanced with $T$ and $\rho$.
They are already comparable to the $\beta^+$ decay rates at $rp$
conditions for the WP nuclei. For more exotic isotopes the rates
are dominated by $\beta^+$ decay, while for more stable isotopes
they are dominated by $cEC$. 

\acknowledgments

This work was supported by Ministerio de Ciencia e Innovaci\'on
(Spain) under Contract No.~FIS2008--01301.

\end{document}